\documentstyle[12pt]{article}
\begin{document}
{\it Cavendish Preprint HEP 93/1 \\
     Adelaide University ADP-93-205/T124 \\
     11th May 1993}
\vspace{1.0cm}
\begin{center}
{\huge \bf Mapping the $x$ dependence of the axial anomaly in \\
polarised deep inelastic scattering  \\}
\vspace{3ex}
\vspace{3ex}
{\large \bf S. D. Bass$^1$ and A. W. Thomas$^2$ \\
\vspace{3ex}
{\it $^{1}$HEP Group, Cavendish Laboratory, \\
University of Cambridge, Madingley Road, Cambridge, CB3 0HE,
England \\}
\vspace{3ex}
{\it $^{2}$Department of Physics and Mathematical Physics,
University of Adelaide, \\
Adelaide, SA 5005, Australia \\ }       }

\vspace{3ex}
{\large \bf Abstract \\}
\end{center}
\vspace{3ex}

{\large
We discuss the role of the U(1) axial anomaly in the spin
structure functions of the nucleon, with particular emphasis on how
one might determine its $x$ dependence in present and future
deep inelastic scattering experiments.
We focus on the C-odd spin structure function $g_3$ and also the
deuteron structure function $g_1^d$.}

\vspace{3.0cm}

\pagebreak

In the past few years a great deal of attention has focussed on the QCD
improved parton model as a result of the EMC spin effect
(or proton ``spin crisis").
The European Muon collaboration (EMC [1]) extended the earlier SLAC
measurement [2] of the
structure function $g_1(x, Q^2)$ of the polarised proton to smaller x
and hence improved the accuracy with which the first moment was
determined.
In the naive parton model $g_1$ is written as :
$$
g_{1}(x) = {1 \over 2} \sum_{q} {\rm e_{q}}^{2} \Delta q(x)
\eqno(1)
$$
where
$$
\Delta q(x) = (q^{\uparrow} + {\overline q}^{\uparrow})(x) -
			(q^{\downarrow} + {\overline q}^{\downarrow})(x)
\eqno(2)
$$
is the polarised quark distribution.
It is helpful to rewrite $g_1(x)$ in terms of the SU(3) flavour
combinations:
$\Delta u(x) - \Delta d(x)$,
$\Delta u(x) + \Delta d(x) - 2 \Delta s(x)$
and
$\Delta u(x) + \Delta d(x) + \Delta s(x)$.
Then the first moment of the flavour singlet piece
is related to the fraction of the proton's spin which is carried by its
quarks.
After a smooth Regge extrapolation of their data ($g_1 \sim x^{-0.12}$)
EMC determined this quantity to be [1]
$$
\Delta u + \Delta d + \Delta s = 0.120 \pm 0.094 (stat.) \pm 0.138 (syst.)
\eqno(3)
$$
which is consistent with zero and two standard deviations
from the Ellis-Jaffe hypothesis, which says that strange quarks should
not play a significant role.
A detailed discussion of the EMC spin experiment and its theoretical
interpretation may be found in the reviews [3] and [4].

As Veneziano has stressed, the EMC result is a violation of Zweig's rule
in the flavour singlet channel [5]. As this is the only one of the three
SU(3) flavour combinations which can involve the U(1) axial anomaly it
seems highly likely that this might be
the source of the spin effect.
If this is the case, and we strongly
suggest that it is,
it is clearly an important experimental problem to map out the $x$
dependence of the axial anomaly in the inclusive DIS cross section.
In this paper we discuss how important information may be
obtained about this in current and future experiments.
We focus on the C-odd spin dependent structure function
$g_3(x)$ and also the polarised deuteron stucture function $g_1^d (x)$.

In QCD the inclusive deep inelastic process is described by the operator
product expansion (OPE) and the renormalisation group.
The interesting physics of $g_1$ is in the flavour singlet
part, which receives contributions
from both quark and gluon partons, viz.
$$
g_{1}(x, Q^{2}) |_{S} = {1 \over 3} \sqrt {2 \over 3}
\int_x^1 {dz \over z}
\Biggl[ \Delta q_{0}(z, Q^2)
C^{q}_{S} ( {x \over z}, \alpha_s (Q^{2}) )
+ {1 \over \sqrt{6} } \Delta g(z, Q^2)
C^g_S ({x \over z}, \alpha_s (Q^2) ) \Biggr]
\eqno(4)
$$
The C-even, spin dependent quark $\Delta q_0(x, Q^2)$
and gluon $\Delta g(x, Q^2)$ distributions are defined with respect
to the operator product expansion.
Their odd moments project out the target matrix elements
of the renormalised, spin odd, composite operators
$$
2M s_+ (p_+)^{2n} \int^1_0 dx \ x^{2n} \Delta q_{k} (x, Q^2) =
<p,s | [ {\overline q}(0) \gamma_+ \gamma_5 (i D_+)^{2n} {\lambda^k \over 2}
q(0) ]_{Q^2}^{GI} |p,s >_c
\eqno(5)
$$
$$
2M s_+ (p_+)^{2n} \int^1_0 dx \ x^{2n} \Delta g (x, Q^2) =
<p,s | [ {\bf \rm Tr} \ G_{+ \alpha}(0) (iD_+)^{2n-1}
{\tilde G}^{\alpha}_{\ +}(0) ]_{Q^2}^{GI} |p,s >_c
(n \geq 1)
\eqno(6)
$$
It is known from unpolarised DIS experiments that the gluon distribution
is concentrated at small $x$. In polarised DIS
the hard photon scatters from a gluon via a quark-antiquark
pair, described in $C^g (x, \alpha_s)$.
This dissipates the gluon's already small momentum so that
$\Delta g(x, Q^2)$ is relevant to $g_1$ only at small $x$ ($x \leq 0.03$)
[6].
It makes a negligible contribution to the measured
sum rule between $x = 0.01$ and 1, where the three constituent quarks are
expected to dominate.

The clue to understanding the spin effect lies in the
identification of the
axial-vector current (and the higher spin axial tensors in equ. (5))
with spin.
Here we encounter a subtle effect in quantum field theory related to
the axial anomaly.
Classically the axial vector current looks like a gauge invariant
operator.
The quark field operator transforms as
$$
q(x) \rightarrow U(x) q(x)
$$
and
$$
{\overline {q}}(x) \gamma_{\mu} \gamma_{5} \rightarrow
  {\overline {q}}(x) \gamma_{\mu} \gamma_{5} U^{\dagger}(x)
\eqno(7)
$$
under a given gauge transformation $U$.
On the other hand, in quantum field theory
the axial vector current operator is not just
${\overline q}(0) \gamma_{\mu} \gamma_{5}$
multiplied by $q(0)$. It is a composite operator which has to be
renormalised and there are extra
divergences which are intrinsic to the operator itself.
It turns out that one cannot renormalise the axial tensor operators
in a gauge invariant way so that they describe spin at the same time.
In general,
for a given choice of renormalisation prescription $R$, the renormalised
axial tensor operator
differs from the gauge invariant operator by a multiple of a
gauge-dependent, gluonic counterterm $k_{\mu \mu_1 ... \mu_{2n}}$, viz.
$$
\biggl[
{\overline q}(0) \gamma_{\mu} \gamma_5 D_{\mu_1} ... D_{\mu_{2n}} q(0)
\biggr]^R_{Q^2} =
\biggl[
{\overline q}(0) \gamma_{\mu} \gamma_5 D_{\mu_1} ... D_{\mu_{2n}} q(0)
\biggr]^{GI}_{Q^2}
+ \lambda_{R, n} \biggl[ k_{\mu \mu_1 ... \mu_{2n} } \biggr]_{Q^2}
\eqno(8)
$$
where the coefficients $\lambda_{R, n}$ are fixed by the choice of
renormalisation prescription.
This axial anomaly was discovered for the axial vector current in QED
[7,8].

Not only does the axial anomaly lead to a difference between the
renormalised axial currents which preserve gauge invariance and chiral
symmetry, but in addition the gauge invariant axial current is scale
dependent (in this case the scale is $Q^2$). (The anomalous dimension of
the first moment, $\Delta q_{0}$, was first calculated in QCD by Kodaira
[9].)
This means that one cannot derive the generators of the spin algebra
SU(2) from it.
It follows [10] that
the gauge invariant axial-vector current and the higher spin
operators which appear in equ. (5) do not describe a distribution of
quark spin in the proton.
One can construct a distribution which does measure spin.
It differs from the
physical distribution $\Delta q (x, Q^2)$ which
is measured in deep inelastic scattering by
by a gauge dependent gluonic term related to the $k_{\mu \mu_1 ... \mu_{2n}}$
in equ.(8) - the anomaly.
(A technical discussion of these issues is given in ref.[10].)
In other words,
one can say that {\it the gauge symmetry screens the spin of the quarks.}

We now compare $g_1$ with the other structure functions measured in
deep inelastic scattering.
The axial anomaly is not relevant to the unpolarised quark distributions,
which are described in OPE language by the operators
$ {\overline q}(0) \gamma_+ (iD_+)^{n} q(0)$.
Nor is it relevant to $g_3$, which is the polarised
version of $F_3$.
Since $g_3$ is odd under charge conjugation, and gluons are C-even,
it can have no anomalous gluonic contribution.
This means that it does make sense to talk about $F_1$, $F_3$ and $g_3$
in terms of quarks with explicit spin degrees of freedom - the clash of
symmetry between gauge invariance and spin does not manifest itself in
these structure functions.

In order to deal with $g_1$ we should modify the parton spin
identification in equ. (2) by writing
the gauge invariant distribution as
$$
q^{\uparrow}_{GI}(x, Q^2) = ( q^{\uparrow}_S + {1 \over 4} \kappa ) (x, Q^2)
\eqno(9a)
$$
$$
q^{\downarrow}_{GI}(x, Q^2) = ( q^{\downarrow}_S - {1 \over 4} \kappa ) (x,
Q^2)
\eqno(9b)
$$
for both quarks and anti-quarks.
Here $\kappa (x, Q^2)$ denotes the anomaly and the subscript $S$ denotes
the gauge-dependent spin distribution.
The $\kappa$ distribution
appears only in the treatment of $g_1$ in deep inelastic scattering.
Since the anomaly is independent of quark flavour the same $\kappa$
distribution is relevant to each of $u, d, s, c, ..$.
It varies only according
to the ``spin" and not the charge
or flavour.
Because $\kappa$ is flavour independent it will induce some OZI violation
wherever it plays a role.
This is the likely source of the EMC spin effect.

If we substitute equs.(9) into the OPE expression for $g_1$ (viz. equ.(4))
 it is easy to see that both the spin and anomalous components of
$\Delta q (x, Q^2)$ couple to the
hard photon in exactly the same way as one expects of a quark
(via $C^q(x,\alpha_s)$).
Physically, this means that there is a
new local interaction between the hard photon and a gluonic component in
the proton, which must be included in the
parton model [11,12].
This is despite the fact that the glue does not carry electric charge !

It is clearly an important problem to map out the $x$ dependence of
the anomaly in $g_1$.
There are some theoretical clues which may provide insight into what
to expect.
We first consider the trace anomaly.
In his classic paper on anomalies Gribov [13] showed that the trace and
axial anomalies are intimately related
at the level of operator renormalisation.
Thus, one might expect the axial anomaly to play a role over the same $x$
range as the trace anomaly.
It is well known [14]
that the proton mass
is determined by the non-vanishing trace of the energy momentum tensor,
that is the scale (or trace) anomaly.
Clearly, gluons are important here: in a free quark model with no
glue the ``proton" would be three
massless, unconfined quarks with total mass zero !
In a semi-classical quark model (say the MIT bag model) this gluonic
component appears as the infinite confining potential well in which the
quarks live.
When we say that the three valence quarks are at large $x$ (viz. that
they carry a lot of the proton's momentum) we should
remember that this proton mass or momentum is generated via the trace
anomaly.
In this way, it makes sense to think of the trace anomaly as a
large $x$ effect.

Let us suppose that the axial anomaly is also manifest over a complete
range of $x$.
(It is an intrinsic part of the physical spin dependent quark distribution.)
In this case, there would be no reason to expect $g_1$ and $g_3$
to be the same at large $x$.
This is in contrast with the unpolarised structure functions, where
the unpolarised C-even and C-odd quark distributions are found to be
identical within (small) errors at large $x \geq 0.2$.
(The large $x$ region is described by the three constituent quarks, whereas
the sea of quark anti-quark excitations and glue is a small $x$
effect.)
In terms of polarisation measurements the C-odd
structure function $g_3$ does measure quark spin since it is anomaly free.
(The traditional quark model predictions should be
reliable here.)
In $g_1$ the spin is screened via the anomaly.
Thus any significant
difference between $g_1$ and $g_3$ at large $x$
would be directly related to the anomaly.
This does not mean that we would isolate $\kappa$ per se by comparing $g_1$
with $g_3$ ($\kappa(x, Q^2)$ is gauge dependent).
Rather, any finite difference at large $x$ would signify a Zweig's rule
breaking effect
in the large $x$ bins and this is the anomaly.

Unfortunately, the cross section for DIS with a neutrino beam and proton
target is very small - enough to
make direct measurements of $g_3$ impracticable at the present time.
However,
if one assumes that the quark fragmentation functions are spin independent
then
it may be possible to extract the C-odd distribution from the
$g_1$ measurements by detecting fast pions from among the final state hadrons
[15].
This experiment is planned by the HERMES collaboration at HERA [16].

Important information about the $x$ dependence of the axial anomaly
in polarised deep inelastic scattering will also come from
measurements of $g_1^n (x, Q^2)$.
The axial anomaly occurs only in the flavour singlet part of $g_1$ and
therefore it will be present equally in
$g_1^p$ and $g_1^n$ as a function of $x$.
If the anomaly acts to screen the quark spin at large $x$
in $g_1^p$
it follows that the same should be true in $g_1^n$.
The combination appearing in the Bjorken sum-rule $(g_1^p-g_1^n)$
has no flavour singlet component and is anomaly free.
On the other hand, the flavour singlet component is enhanced in the
deuteron structure function
$g_1^d ={1\over2} (g_1^p + g_1^n)$,
which has no isotriplet piece $\Delta q_3 (x, Q^2)$.
Thus the deuteron structure function $g_1^d$ is an ideal place to test
model predictions about how the anomaly should contribute in the nucleon
structure function $g_1 (x,Q^2)$.

The usual quark model calculations, which do not
include the anomaly,
suggest that $g_1^n$ will change sign and become small and positive
at large $x$ [17-20].
To the extent that these models do not include any OZI violation,
a large $x$ anomaly would tend to render $g_1^n$ negative at large $x$.
As a specific example we consider
the quark model calculation of the structure functions
which was developed by the Adelaide group [18-20] following earlier
work by Jaffe, Ross and others [21].
These calculations provide reasonable agreement with the unpolarised
structure function data.
In the form used by these authors, the bag model has not yet been
extended to satisfy the U(1) chiral Ward identity.
That is, it does not include an OZI violation induced by the anomaly.
However it does seem reasonable that these model calculations for $g_1$
might describe
$g_3 (x)$ at large $x$ - i.e. correspond to a world without the OZI violations
due to the anomaly.

The bag model prediction for $g_1 (x)$ overestimates the data
throughout all of the large $x$ region [19,20].
Hence, it is tempting to associate the difference between the
model results and the data as that associated with the anomaly.
This is illustrated in Fig. 1a.
Here we show the EMC and earlier SLAC data for $x g_1^p$ together with the
naive bag model expectation
(the solid line) corresponding to a bag radius $R = 0.8$fm, which we take
from Schreiber et al. [19].
The dashed curve is the result of adding a purely phenomenological
term to fit the data [22].
We will associate this phenomenological term with the OZI violation
missing in the naive bag model.
In this case, the same flavour singlet correction should therefore be applied
to the neutron
and this is shown in Fig. 1b.
Here the solid curve is the naive bag (no OZI violation) prediction for
$x g_1^n$ (again at $Q^2 = 10$GeV$^2$ for a bag radius $R=0.8$fm) and
the dashed curve is obtained by adding the same correction that was
applied for the proton. Clearly when this correction is included
we find that the neutron spin-dependent structure
function becomes negative at large $x$.
In Fig. 1c we show the naive bag (solid) and anomaly corrected (dashed)
predictions for $x g_1^d$.
(For the present purposes we make the simple approximation that
$ g_1^d = ( g_1^p + g_1^n )/2$ ,thus ignoring corrections due to
shadowing,Fermi-motion and the D-state probability of the deuteron.
These are expected to be important at the few-percent level [23] - well below
the present experimental accuracy.)
The corrected curve is in good agreement with the recent SMC measurement
of the deuteron spin structure function $x g_1^d(x,Q^2)$ [24].

We repeat this analysis in Figs. 2a-2c for a bag radius of $R=0.6$fm.
The bag model calculation of $g_1$ is taken from ref.[20], while the
dashed curve is again the bag result supplemented by a purely
phenomenological term [25].
Again the bag model supplemented by the anomaly term
tends to favour a negative sign for $g_1^n$ at large $x$,
and there is reasonable agreement with the SMC data for $x g_1^d$.

The possibility that $g_1^n$ might be negative at large $x$
was also raised by Benesh and Miller [26].
These authors were working within Jaffe's hypothesis [27] that
there is a large change in $\Delta q_0$ between some low scale, $\mu ^2$,
typical of a quark model and the higher values of $Q^2$ where scaling sets in.
They assumed that Jaffe's hypothesis for the first moment might also
be true for all higher moments and simply set $\Delta q_0(x,Q^2)$ to zero
above about 1GeV$^2$.
As we have shown elsewhere [28], this scenario would also imply
a sudden
and dramatic jump in $\int_0^1 dx g_1^p (x, Q^2)$ as we go through the
charm threshold.
At the present time, there is no experimental evidence for such a jump,
but more precise data over a wide range of $x$ through the charm
threshold would be most helpful.

In summary, we have discussed how one could map out the $x$ dependence of
the axial anomaly in $g_1$.
If the anomaly is a large $x$ effect then it can be isolated as a
finite difference between $g_1$ and $g_3$ in the large $x$ bins. If it
is purely a small $x$ effect the anomaly would be lost among the sea
and gluon
distributions which dominate the data at small $x$ (say $\leq 0.1$).
We stress that the comparison with $g_3$ is the only definitive experimental
test of whether the anomaly is a large or a small $x$ effect.
Certainly, it is an intrinsic part of the spin dependent quark distribution
and there is no good theoretical reason to believe that it is confined to
small $x$.
We strongly urge that consideration be given to the challenging
experimental
problem of how to measure $g_3$.
In the interim it would be very useful to obtain more data (with reduced
errors) on the deuteron spin structure function $g_1^d$.
This deuteron data will help constrain theoretical models of the structure
functions.

\vspace{1.0cm}

This work was supported in part by the Australian Research Council.

\pagebreak

\begin{center}{\bf References}\end{center}
\vspace{1.0cm}
\begin{enumerate}
\item
J. Ashman et al., Phys. Lett. B206 (1988) 364, Nucl. Phys. B328 (1990) 1.
\item
G. Baum et al., Phys. Rev. Lett. 51 (1983) 1135.
\item
R. Windmolders, Int. J. Mod. Phys. A7 (1992) 639.
\item
S. D. Bass and A. W. Thomas, Cavendish Preprint HEP 92/5 (1992),
to appear in J. Phys. G.
\item
G. Veneziano, Okubofest lecture, CERN preprint TH-5840/90 (1990).
\item
S. D. Bass and A. W. Thomas, Nucl. Phys. A527 (1991) 519c;
J. Ellis, M. Karliner and C. Sachrajda, Phys. Lett. B231 (1989) 497;
S. D. Bass, N. N. Nikolaev and A. W. Thomas, Adelaide preprint
ADP-90-133/T80 (1990).
\item
J. S. Bell and R. Jackiw, Nuovo Cimento 51A (1969) 47.
\item
S. L. Adler, Phys. Rev. 177 (1969) 2426.
\item
J. Kodaira, Nucl. Phys. B165 (1980) 129.
\item
S. D. Bass, Cavendish preprints HEP 92/11 (1992) and 93/2 (1993).
\item
S. D. Bass, Zeit. Phys. C55 (1992) 653.
\item
V. N. Gribov, remark at SLAC Lepton Photon Symposium, see proceedings
page 59, ed. M. Riordan, World Scientific (1990).
\item
V. N. Gribov, Budapest preprint KFKI-1981-66 (1981).
\item
M. A. Shifman, A. I. Vainshtein and V. I. Zakharov,
Nucl. Phys. B147 (1979) 385, 448;
R. L. Jaffe and A. Manohar, Nucl. Phys. B337 (1990) 509.
\item
L. L. Frankfurt et al., Phys. Lett. B320 (1989) 141.
\item
HERMES Proposal, K. Coulter et al., DESY/PRC 90-1 (1990);
M. Veltri et al., Proc. Physics at HERA, Vol. 1, 447 (1991).
\item
R. Carlitz and J. Kaur, Phys. Rev. Lett. 38 (1977) 673;
A. Schafer, Phys. Lett. B208 (1988) 175.
\item
F. E. Close and A. W. Thomas, Phys. Lett. B212 (1988) 227.
\item
A. W. Schreiber, A. W. Thomas and J. T. Londergan, Phys. Rev. D42 (1990) 2226.
\item
A. W. Schreiber, A. I. Signal and A. W. Thomas, Phys. Rev. D44 (1991) 2653.
\item
R. L. Jaffe and G. G. Ross, Phys. Lett. B93 (1980) 313;
R. L. Jaffe, Nucl. Phys. B229 (1983) 205.
\item
The phenomenological term is taken to be
$ -0.08 x^{0.5} (1-x)^2$ for $x\leq 0.45$, while beyond this
 it is cut-off by multiplying by $\exp (-35(x-0.45)^2)$.
\item
R. M. Woloshyn, Nucl. Phys. A496 (1989) 749;
B. Badelek and J. Kwiecinski, Nucl. Phys. B370 (1992) 278;
W. Melnitchouk and A. W. Thomas, Phys. Rev. D47 (1993) 3783.
\item
The SMC Collaboration, Phys. Lett. B302 (1993) 533.
\item
For $R=0.6fm$ the phenomenological term is taken to be
$ -0.07 x^{0.4} (1-x)^3$ for $x \leq 0.45$, which
is multiplied by the large $x$ cut-off
$\exp(-5 (x-0.45)^2)$ for $x\geq 0.45$.
\item
C. J. Benesh and G. A. Miller, Phys. Lett. B222 (1989) 476.
\item
R. L. Jaffe, Phys. Lett. B193 (1987) 101.
\item
S. D. Bass and A. W. Thomas, Phys. Lett. B293 (1992) 457.
\end{enumerate}

\pagebreak

\begin{center} {\bf Figures} \end{center}
\begin{enumerate}
\item
{\bf Figs. 1a-1c:}

Fig.1a shows the world data for $x g_1^p$ [1,2] together
with the bag model expectation
[19] (solid curve), for a bag radius $R=0.8$fm.
The dashed curve is obtained by adding a phenomenological
term to fit the data.
Figs.1b and 1c show
the naive bag (solid curve) and OZI corrected (dashed curve) prediction
for $xg_1^n$ and $xg_1^d$ respectively, together with the
SMC data for $xg_1^d$ [23].
\item
{\bf Figs. 2a-c:}

Fig.2a shows the world data for $x g_1^p$ [1,2]
together with the bag model expectation
[20] (solid line), for a bag radius $R=0.6$fm.
We fit the data by adding a phenomenological term (dashed curve).
Figs.2b and 2c show
the naive bag (solid curve) and OZI corrected (dashed curve) prediction
for $xg_1^n$ and $xg_1^d$ respectively, together with
the SMC data for $xg_1^d$ [23].
\end{enumerate}
\end{document}